# Measurement of anharmonicity of phonons in the negative thermal expansion compound $Zn(CN)_2$ by high pressure inelastic neutron scattering

R. Mittal[1,2], S. L. Chaplot[2] and H. Schober[3]

[1]*Juelich Centre for Neutron Science, IFF, Forschungszentrum Juelich, Outstation at FRM II, Lichtenbergstr. 1, D-85747 Garching, Germany*

[2]*Solid State Physics Division, Bhabha Atomic Research Centre, Trombay, Mumbai 400 085, India*

[3]*Institut Laue-Langevin, BP 156, 38042 Grenoble Cedex 9, France*

$Zn(CN)_2$ is known to have an isotropic negative thermal expansion (NTE) coefficient (about $-51 \times 10^{-6}$ $K^{-1}$) over 10-370 K that is twice as large as that of $ZrW_2O_8$. We have measured the pressure dependence of the phonon spectra up to 30 meV from a polycrystalline sample of $Zn(CN)_2$ at pressures of 0, 0.3, 1.9 and 2.8 kbar at temperatures of 165 and 225 K. The measurements enabled us to estimate the energy dependence of the ratios $\frac{\Gamma_i}{B}$ ($\Gamma_i$ are Grüneisen parameters as a function of phonon energy $E_i$ at ambient pressure and B is the bulk modulus), which reflect the anharmonicity of phonons. We conclude that the phonon modes of low energy below 15 meV play an important role in the understanding of the NTE behavior in $Zn(CN)_2$ and the measured anharmonicity can quantitatively explain the NTE.

Studies of materials exhibiting unusual properties like negative thermal expansion (NTE) are of interest due to both their fundamental scientific importance and potential applications [1-3]. Composites with tailored thermal expansion are useful in high precision applications. The NTE materials find important scientific as well as commercial interest as they can be used to make composites with nearly zero thermal expansion coefficients by compensating with the usual positive thermal expansion coefficient of other materials. Besides oxide-based framework materials, NTE behavior has been observed in molecular framework materials containing linear diatomic bridges such as the cyanide anion. For example, $Zn(CN)_2$ is reported [2] to have a NTE coefficient ($-51 \times 10^{-6}$ $K^{-1}$) twice as large as that of $ZrW_2O_8$ [1].

Structural studies [4] on $Zn(CN)_2$ show that two different models having cubic symmetry with space group *Pn3m* (disordered model) and *P43m* (ordered model) fit equally well to the diffraction data and give similar agreement factors. The ordered structure consists of a $ZnC_4$ tetrahedron (at the centre of the cell) linked to four neighboring $ZnN_4$ tetrahedra (at the corners of the cell) with CN bonds along the body-diagonals. On the other hand, in the disordered structure CN bonds are orientationally disordered or flipped randomly such that both C and N sites are occupied by these atoms with a fractional occupancy of 0.5. The refinement of the neutron diffraction data [4] at 14 K with the ordered model gives unreasonable atomic displacement factors for C (0.0013 $Å^2$) and N (0.0152 $Å^2$) in comparison to the isotropic value of 0.0089 $Å^2$ for C, N atoms for the disordered model.

$Zn(CN)_2$ is fundamentally different from other NTE materials ($ZrW_2O_8$ etc) in that the structure of $Zn(CN)_2$ is based on a framework of Metal-Cyanide-Metal (M-CN-M) rather than Metal-Oxygen-Metal (M–O–M) linkages. Atomic pair distribution function (PDF) analysis [5] of high energy X-ray scattering data show an increase of the average transverse thermal amplitude of the bridging C/N atoms away from the body diagonal with heating from 100 to 400 K. This increase of the thermal amplitude of the bridging atoms is believed to be at the origin of the NTE behavior in $Zn(CN)_2$.

Normally solids become less compressible at high pressure with a positive pressure dependence of the bulk modulus. However, high pressure neutron diffraction measurements [6] carried out for $Zn(CN)_2$ at 300 K show that it becomes more compressible with an increase of pressure (Bulk modulus B= 34.19 GPa, pressure derivative of B, B'= -6.0). X-ray diffraction measurement [7] carried out at 300 K gives a bulk modulus value of 25 $\pm$ 11 GPa. Synchrotron x-ray diffraction experiments [8] under hydrostatic conditions show that $Zn(CN)_2$ transforms from a cubic to an orthorhombic phase at about 1

GPa. The orthorhombic phase transforms to another cubic-II phase at about 2 GPa followed by amorphization of Zn(CN)$_2$ at about 11 GPa. Two independent ab-initio calculations [7,9] of the phonon spectra and Grüneisen parameters have been reported recently that show large differences. The calculated bulk modulus at 0 K from ab-initio calculations [7,9] is much larger (59 GPa from Ref. [9] and 88 GPa from Ref. [7]) in comparison of the experimental values [6,7] at 300 K.

The study of the mechanism of NTE essentially involves [10-19] the identification of anharmonic phonons and their softening on compression of the crystal. In thermodynamics literature a Grüneisen parameter is defines in terms of the ratio of the thermal expansion and the specific heat [20]. However, here we investigate detailed contributions of the anharmonicity of individual phonons in the crystal. The anharmonicity of phonon modes as reflected in their individual mode Grüneisen parameters, $\Gamma_i = -\frac{V}{E_i}\left(\frac{\partial E_i}{\partial V}\right)$ (where V and $E_i$ are unit cell volume and phonon energy respectively) are directly responsible for thermal expansion in a material; it becomes vitally important to study them. In the quasiharmonic approximation, the volume thermal expansion coefficient [ $\alpha_V = \frac{1}{V}\sum_i \frac{\partial \ln E_i}{\partial P} C_{Vi}(T)$ , where $\frac{\partial \ln E_i}{\partial P}\left(=\frac{\Gamma_i}{B}\right)$, $B = -V\frac{\partial P}{\partial V}$, $C_{Vi}$ is the specific heat, P is pressure and B is bulk modulus] can be determined from the pressure dependence of phonon spectra. Earlier we carried out high pressure inelastic neutron scattering experiments [16,17] on polycrystalline samples of cubic ZrW$_2$O$_8$ and ZrMo$_2$O$_8$ using the IN6 spectrometer at the Institut Laue Langevin (ILL), France. It has been shown that low-energy phonon modes play an important role in understanding the NTE behavior [10-12,16,17]. Raman spectroscopy at high pressure [7] has been used to experimentally determine Grüneisen parameters values of Zn(CN)$_2$ for modes above 200 cm$^{-1}$ (25 meV). The low-energy part of the phonon spectra has been measured [21] using the PRISMA spectrometer at ISIS.

A single crystal of Zn(CN)$_2$ is not available for the measurement of the pressure dependence of the phonon frequencies. Using a polycrystalline sample, we have experimentally determined $\frac{\partial \ln E_i}{\partial P}\left(=\frac{\Gamma_i}{B}\right)$ and its temperature dependence using high-pressure inelastic neutron scattering experiments at pressures up to 2.8 kbar. These measurements with the polycrystalline sample allow us to derive averages over all directions of the phonon wave-vectors. The temperature dependence is important in the quantitative calculation of the thermal expansion coefficient from phonon data.

Zn(CN)$_2$ (~98.0% pure) polycrystalline sample was obtained from Aldirch, USA. The high-pressure inelastic neutron scattering experiments for Zn(CN)$_2$ were carried out using the IN6 spectrometer at the ILL. An incident neutron energy of 3.12 meV was chosen for the measurements which led to a resolution of 80 μeV at the elastic line. About 7 grams of Zn(CN)$_2$ sample was compressed using argon gas in a pressure cell available at ILL. The use of argon gas as a pressure transmitting medium allowed us to perform the measurements above its critical point at 160 K. Synchrotron x-ray diffraction [8] does not show any phase transition for Zn(CN)$_2$ up to 1 GPa. We have measured the inelastic neutron scattering spectrum from a polycrystalline sample of Zn(CN)$_2$ at ambient pressure, 0.3 kbar, 1.9 kbar and 2.8 kbar, at 165 K and 225 K. The data were taken over a range in scattering angle from 10$^o$ to 113$^o$. The inelastic neutron scattering signal is corrected for the contributions from argon at the respective pressures and for the empty cell. The incoherent approximation [22] was used in the data analysis. The data were suitably averaged over the angular range of scattering using the available software package at ILL to obtain the neutron-cross-section weighted phonon density of states.

The measured spectrum from Zn(CN)$_2$ as a function of temperature at various pressures is shown in Fig. 1 for energy transfers up to 30 meV. The ambient pressure measurements are in agreement with the previous measurements [21]. Ab-initio phonon calculation [9] shows that the spectral weight up to 30 meV is 46.67 %, which corresponds to 14 phonon modes out of a total of 30 modes in cubic Zn(CN)$_2$ per unit cell. The striking feature in the spectrum is the very strong low energy peak at 2 meV. There is a continuous spectrum of excitations in the measured energy transfer range, with maxima at 8 meV and 18 meV. The low energy peak at about 2 meV in the phonon spectra of Zn(CN)$_2$ (Fig. 1) appears to be from a flat transverse acoustic mode [9]. The band around 8 meV is more likely due to hybridization of acoustic and optic modes. The Grüneisen parameters of these mode are needed for understanding the NTE behavior of Zn(CN)$_2$.

The $\frac{\partial \ln E_i}{\partial P}\left(=\frac{\Gamma_i}{B}\right)$ values for phonons of energy $E_i$ have been obtained at 165 K and 225 K [Fig. 2] using the cumulative distributions of the density of states. The modes up to 15 meV show negative $\frac{\Gamma_i}{B}$, with the low-frequency modes showing the largest negative $\frac{\Gamma_i}{B}$. The $\frac{\Gamma_i}{B}$ values calculated using the ab-initio calculations are also shown in Fig. 2. These values compare very well with our experimental data for energies above 5 meV. The estimated $\frac{\Gamma_i}{B}$ values at 165 K and 225 K are within the experimental error bars. The $\frac{\Gamma_i}{B}$ (Fig. 2) and experimental phonon spectra, $g^{(n)}(E)$ (Fig. 1) at 165 K have been used for the calculation of $\alpha_V$ (Fig.3(a)) using the relation for $\alpha_V$ given above. The comparison of the volume thermal expansion derived from the phonon data and diffraction data [2] is shown in Fig. 3(b), which shows a good agreement between them. Thus the anharmonicities of low phonon modes are sufficient to account for the negative thermal expansion coefficient of $Zn(CN)_2$.

The Grüneisen parameters derived from the present experiment cannot be compared with those derived from the Raman and infrared data [7] since these measurements report observation of phonon modes in $Zn(CN)_2$ only above 200 cm$^{-1}$ (~25 meV). First principles ab-initio phonon calculations [9] show that nearly dispersion-less transverse acoustic modes do appear in the energy range of 2-4 meV and have negative Grüneisen parameters of about -7. Our experimental data show that for these modes $\Gamma_i$ values lie in between -14 and -9. The negative volume thermal expansion coefficient from ab-initio calculations [9] at 5 K is $-12 \times 10^{-6}$ K$^{-1}$, while the NTE coefficients calculated from our phonon data are shown in Fig. 3(a).

The pair distribution function (PDF) analysis [5] of high-energy X-ray scattering data indicate an increase in the average transverse vibrational amplitude of C/N bridging atoms with increasing temperature, which may provide a mechanism for the NTE in $Zn(CN)_2$. The experimental $\frac{\Gamma_i}{B}$ values (Fig. 3(a)) at 165 K and 225 K have been used for the estimation of the contribution of various phonons to the thermal expansion (Fig. 4) as a function of phonon energy at 165 K and 225 K. Our analysis shows (Fig. 4) that the maximum negative contribution to $\alpha_V$ is from the low-energy transverse acoustic modes of energy of about 2.5 meV, which is consistent with the PDF analysis [5].

In summary, we have identified the low-energy transverse acoustic phonon modes responsible for NTE in $Zn(CN)_2$ from high-pressure inelastic-neutron scattering experiments. The measurements show that $\frac{\Gamma_i}{B}$ values are nearly the same at 165 K and 225 K. The thermal expansion coefficient derived from the phonon data is in good agreement with that obtained from diffraction measurements.

FIG. 1. The experimental phonon spectra, $g^{(n)}(E)$ for $Zn(CN)_2$ as a function of pressure at fixed temperatures of 165 K and 225 K: ambient pressure (full line), 0.3 kbar (dotted line), 1.9 kbar (dashed line), and 2.8 kbar (dash-dotted line).

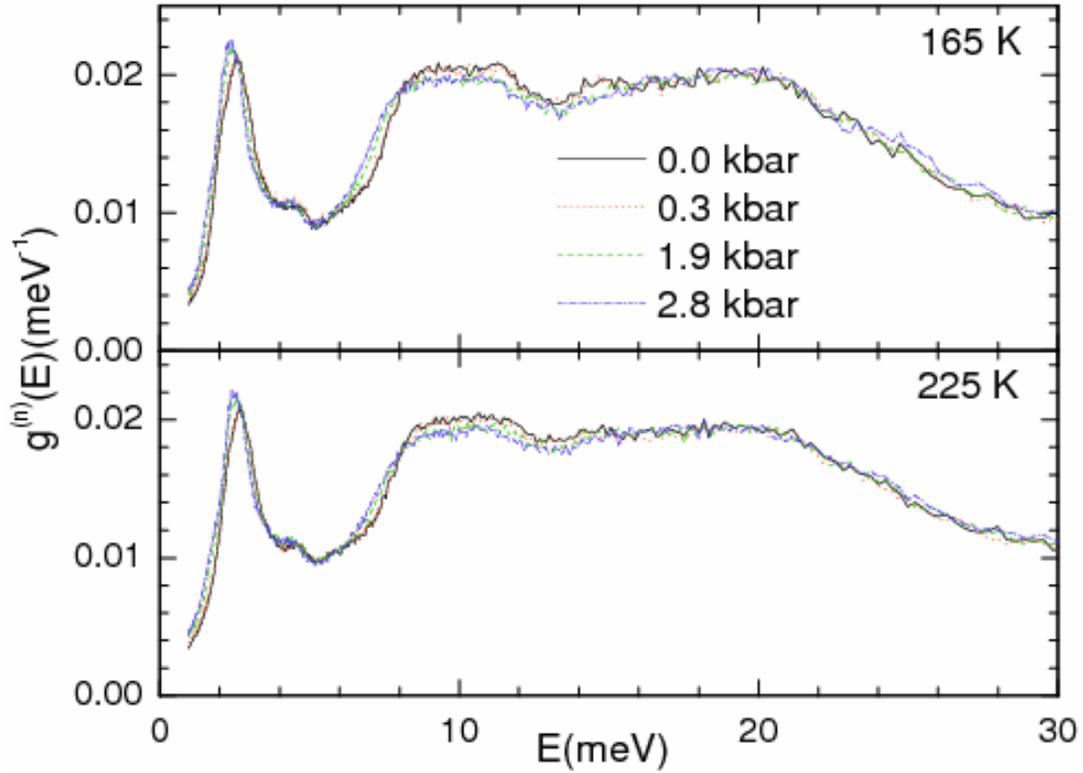

FIG. 2. The experimental $\frac{\Gamma_i}{B}$ as a function of phonon energy $E$ (averaged over the whole Brillouin zone). The $\frac{\Gamma_i}{B}$ values at 165 K, and 225 K has been determined using the density of states at $P = 0$ and 2.8 kbar (full line) which represents the average over the whole range in this study. The $\frac{\Gamma_i}{B}$ values derived from ab-initio calculations [9] are shown by open circles.

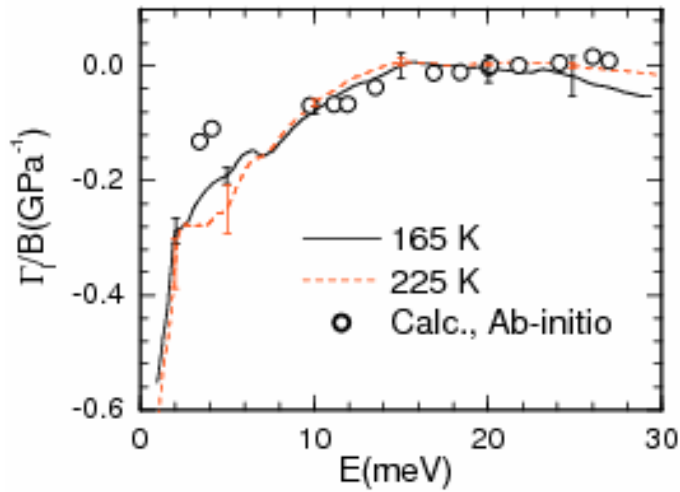

FIG. 3. (a) The volume thermal expansion coefficient ($\alpha_V$) derived from the experimental $\frac{\Gamma_i}{B}$ values at 165 K. (b) The comparison between the volume thermal expansion derived from the present high-pressure inelastic neutron scattering experiment (solid line) and that obtained using x-ray diffraction [2] (open circles).

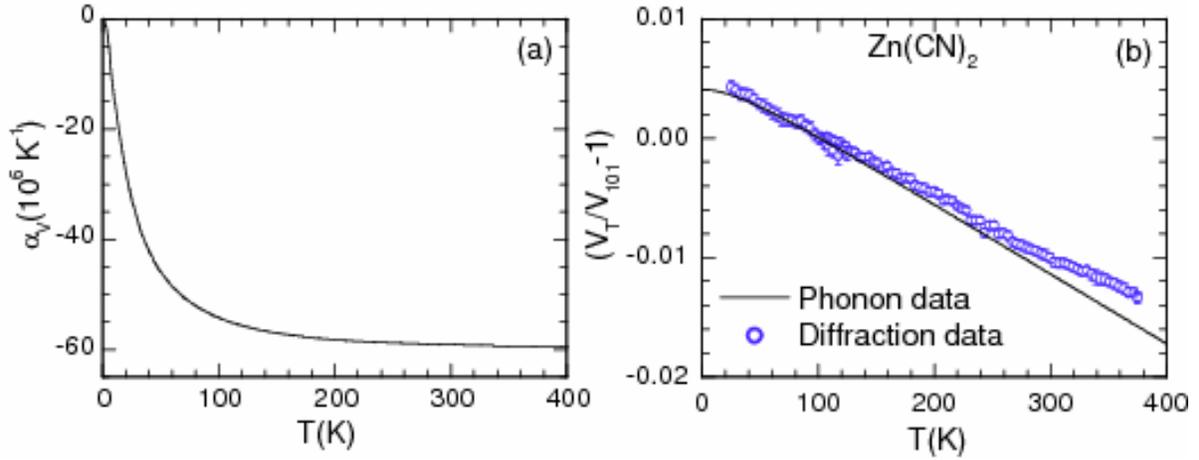

FIG. 4. The contribution of phonons of energy $E$ to the volume thermal expansion coefficient ($\alpha_V$) as a function of $E$ at 165 K and 225 K. The experimental $\frac{\Gamma_i}{B}$ values at 165 K and 225 K have been used for the estimation.

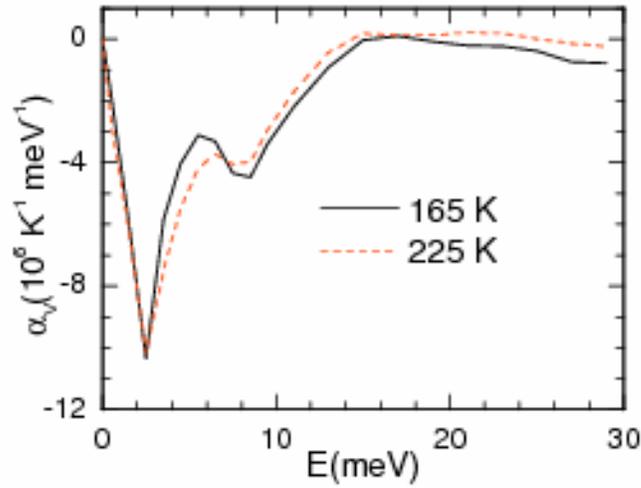